\documentclass[aps,prl,twocolumn]{revtex4}

\usepackage{rotating}     
\usepackage{graphicx}    

\begin{document}

\title{Nonlinear unbalanced Bessel beams in the collapse of Gaussian beams\\ arrested by nonlinear losses}
\author{Miguel A. Porras$^1$, Alberto Parola$^2$}
\affiliation{$^1$Departamento de F\'{\i}sica Aplicada, Universidad Polit\'ecnica de Madrid, Rios Rosas 21, ES-28003, Spain\\
             $^2$CNISM and Department of Physics, University of Insubria, Via Valleggio 11, IT-22100
             Como, Italy}

\begin{abstract}
Collapse of a Gaussian beam in self-focusing Kerr media arrested by nonlinear losses may lead to the spontaneous formation of a
quasi-stationary nonlinear unbalanced Bessel beam with finite energy, which can propagate without significant distortion over
tens of diffraction lengths, and without peak intensity attenuation while the beam power is drastically diminishing.
\end{abstract}

\maketitle

Nonlinear unbalanced Bessel beams (NL-UBB) \cite{porras-UBB-PRL} have been recently described as the only possible type of light
beam that can propagate without distortion and attenuation in a medium with nonlinear mechanisms of dissipation of energy.
Contrary to spatial solitons, NL-UBB transport infinite power, which allows for a continuous transversal energy flux from the
beam periphery towards its nonlinear center, refuelling the power absorbed in this region during propagation.
\cite{porras-UBB-PRL} The idea of NL-UBB has been extended to Bose-Einstein condensates, where similar long-lived structures are
supported by the dissipation provided by three-body inelastic collisions. \cite{victor1}

A spatial dynamics showing the spontaneous transformation of a Bessel beam into a NL-UBB in a Kerr medium with nonlinear losses
(NLL) has been observed in \cite{Pole1}. More importantly, the transformation of an ultrashort pulsed Bessel beam into a pulsed
NL-UBB may take place without the excitation of any temporal dynamics, \cite{Pole2} as the spectral broadening and temporal
splitting that would occur in the Bessel beam in absence of NLL. These NL-UBB stabilized by NLL have indeed been observed in
recent experiments. \cite{Pole2}

The robustness of the NL-UBB suggests that it may act as an attractor of the Kerr dynamics with NLL over a broader space of
input conditions. In this Letter we report on numerical simulations of the propagation of an input Gaussian beam showing the
spontaneous formation of a quasi-stationary light beam with most of relevant features of a NL-UBB. The different regimes of
propagation of a Gaussian beam in a medium with Kerr nonlinearity and NLL were identified in \cite{Poly}. NL-UBBs are seen here
to emerge in case of strong self-focusing stopped by a small amount of NLL. Contrarily to a Bessel beam, the input Gaussian beam
carries finite power. As a consequence, only the nonlinear core of the NL-UBB is formed. Stationarity of the nonlinear core is
seen to be sustained by an energy flux from the beam periphery, as in NL-UBBs. However, since the available power is finite and
decreasing, stationarity lasts after a propagation distance, which is, nevertheless, much larger than the diffraction length
associated to the nonlinear core. Our monochromatic approach is expected to describe also experiments with short pulses insofar
NLL is the mechanism stopping collapse, and can be assumed to quench any significant temporal dynamics afterwards. \cite{Pole2}

\begin{figure}[!b]
\hspace*{-0.6cm}\includegraphics[width=7.8cm]{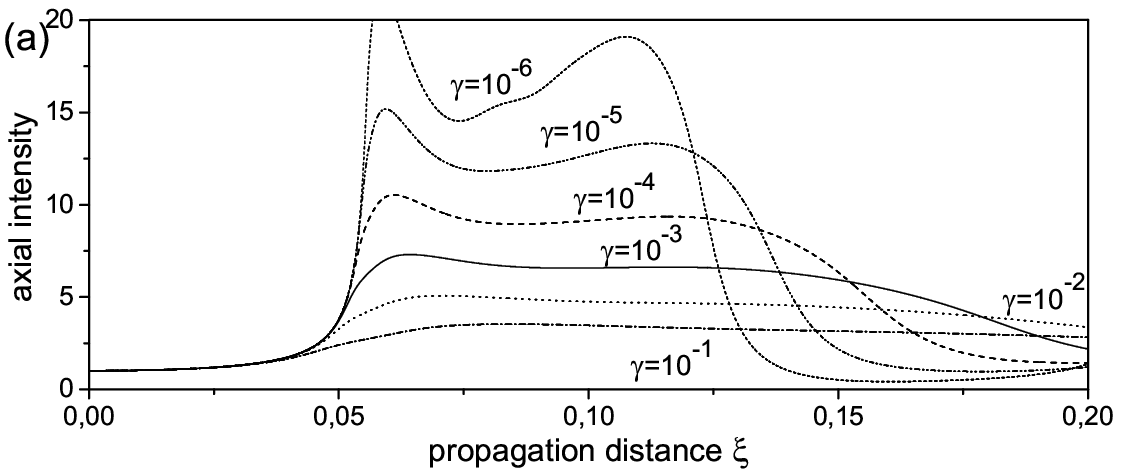}
\includegraphics[width=8.5cm]{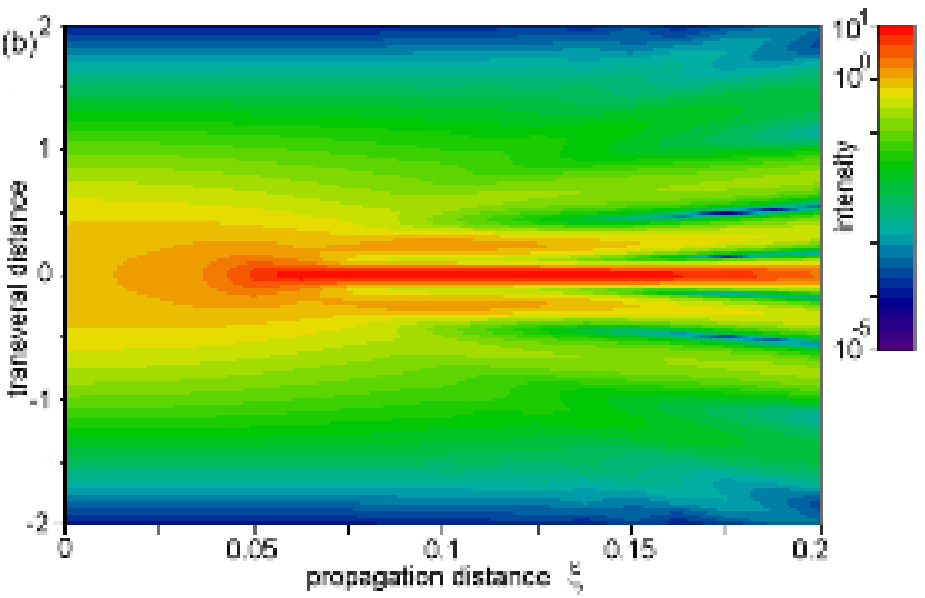}
\caption{\label{axialintensity} (a) On-axis intensity $|\tilde A(0,\xi)|^2$ for input Gaussian with $g=60$ and increasing
$\gamma$ ($K=8$). (b) Intensity $|\tilde A(\rho,\xi)|^2$ for $g=60$ and $\gamma=10^{-3}$ ($K=8$) (in logarithmic scale).}
\end{figure}

The nonlinear Schr\"odinger equation (NLSE), \cite{porras-UBB-PRL,Poly}
\begin{equation}\label{NLSE}
\partial_z A = \frac{i}{2k}\frac{1}{r}\partial_r(r\partial_r A)
+ i \frac{kn_2}{n}|A|^2 A - \frac{\beta^{(K)}}{2}|A|^{2K-2}A\, ,
\end{equation}
is assumed to describe the propagation of a light beam $E=A(r,z)\exp(-i\omega t + i kz)$ with revolution symmetry along the $z$
axis in a medium with Kerr nonlinearity (linear and nonlinear refraction indexes $n$ and $n_2>0$, respectively) and multi-photon
absorption (coefficient $\beta^{(K)}>0$, $K=2,3\dots$). For the incident Gaussian beam $A(r,0)=\sqrt{I_0}\exp(-r^2/s_0^2)$ of
peak intensity $I_0$ and width $s_0$, we introduce dimensionless amplitude $\tilde A=A/\sqrt{I_0}$, radial coordinate
$\rho=r/s_0$, and propagation distance $\xi=z/ks_0^2$, to write (\ref{NLSE}) as
\begin{equation}\label{ANLSE}
\partial_\xi \tilde A = \frac{i}{2}\frac{1}{\rho}\partial_\rho(\rho\partial_\rho \tilde A) + i g|\tilde A|^2 \tilde A - \gamma|\tilde A|^{2K-2}
\tilde A\, ,
\end{equation}
and the initial condition as $\tilde A(\rho,0)=\exp(-\rho^2)$, where $g=k^2n_2I_0s_0^2/n$ and
$\gamma=\beta^{(K)}I_0^{K-1}ks_0^2/2$ measure the strengths of Kerr nonlinearity and of NLL in the input Gaussian beam.

Without NLL, the Gaussian beam collapses if $g> g_0\simeq 3.79$. \cite{Fibich} Figure \ref{axialintensity} (a) displays the peak
intensity $I=|\tilde A(0,\xi)|^2$ along $\xi$, numerically evaluated from (\ref{ANLSE}) with $g=60\gg g_0$, where the strong
collapse is arrested by different amounts of NLL $\gamma$ ($K=8$). Similar trend is observed for other values of $g$, $\gamma$
and $K$ such that the characteristic lengths $\xi_{\rm SF}=1/(2\sqrt{g})$, $\xi_{\rm D}=1/2$, and $\xi_{\rm
NLL}=(2^{K-1}-1)/[2\gamma(K-1)]$ of self-focusing, diffraction and NLL of the input Gaussian (defined as in \cite{Poly}) satisfy
$\xi_{\rm SF}\ll \xi_{\rm D}\ll \xi_{\rm NLL}$. Under these conditions NLL are negligible in the fast self-focusing, becoming
important only in the collapse region. Arrest of collapse by NLL leads to an axial region of sustained high intensity. A rough
estimate of the peak intensities $I$ (in units of the input intensity $I_0$) involved in this region is provided by the simple
formula $I\approx (g/\gamma)^{1/(K-2)}$ (e.g., $I= 19.8, 6.3$ and $2.9$ for $\gamma=10^{-6}, 10^{-3}$ and $10^{-1}$,
respectively) which is obtained, remarkably, by equating the strength $gI$ of the Kerr nonlinearity (for a field of intensity
$I$) to the strength $\gamma I^{K-1}$ of NLL. With increasing $\gamma$, ripples in the peak intensity $I$ disappear, the
intensity then remaining nearly constant for intermediate values of $\gamma$ (between $10^{-4}$ and $10^{-3}$ for $\gamma=60$),
or monotonically, though slowly decreasing for higher $\gamma$.

The transversal profile after collapse features a central spike surrounded by one or more rings, as seen in Fig.
\ref{axialintensity}(b). The width of the central spike can be characterized by $1/\sqrt{2gI}$, and remains therefore constant
upon propagation as long as $I$ is constant. Fig. \ref{axialintensity}(b) corresponds to the case of flattest peak intensity $I$
($\gamma=10^{-3}$), and evidences that the inner portion of the transversal profile, comprising not only the central spike but
also one ring, remains quasi-stationary in the region of flat peak intensity, while outer rings are seen to spread due to
diffraction. The axial length $\simeq 0.05$ of this region is estimated to be about 17 times the diffraction length $s^2/2\simeq
0.003$ associated to the width $s\simeq 0.077$ ($1/e^2$ intensity decay) of the central spike.

\begin{figure}[t]
\begin{center}
\includegraphics[width=8cm]{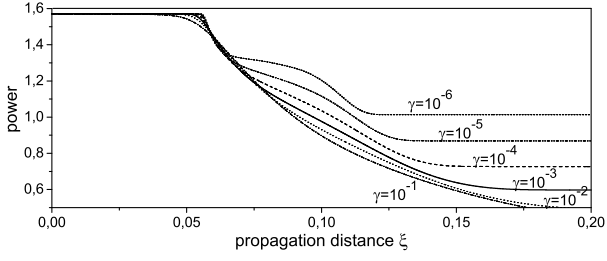}
\end{center}
\caption{\label{power} Beam power $2\pi\int_0^\infty d\rho\rho |\tilde A(\rho,\xi)|^2$ as a function of propagation distance
$\xi$ for $g=60$ and increasing $\gamma$ ($K=8$).}
\end{figure}

Figure (\ref{power}) shows that these properties of stationarity are accompanied, paradoxically, by a drastic diminution of the
beam power due to NLL, which are greatly enhanced in this region of high intensity. Being nonlinear, absorption takes place
mainly in the inner, hot core of the beam, just that remaining stationary. Along the region of flat peak intensity
($\gamma=10^{-3}$), for instance, the power is seen to diminish down to about one half its initial value.

\begin{figure}[b!]
\begin{center}
\includegraphics[width=4cm]{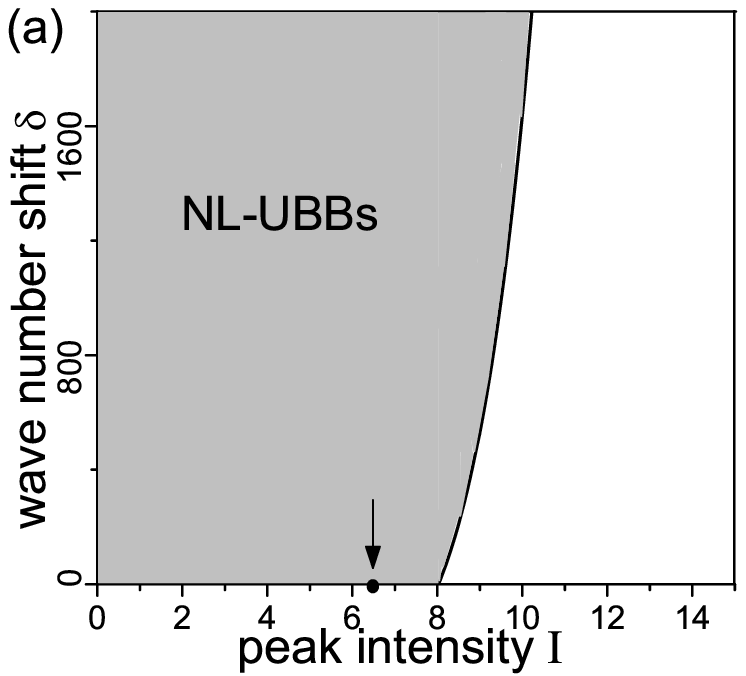}\includegraphics[width=4.3cm]{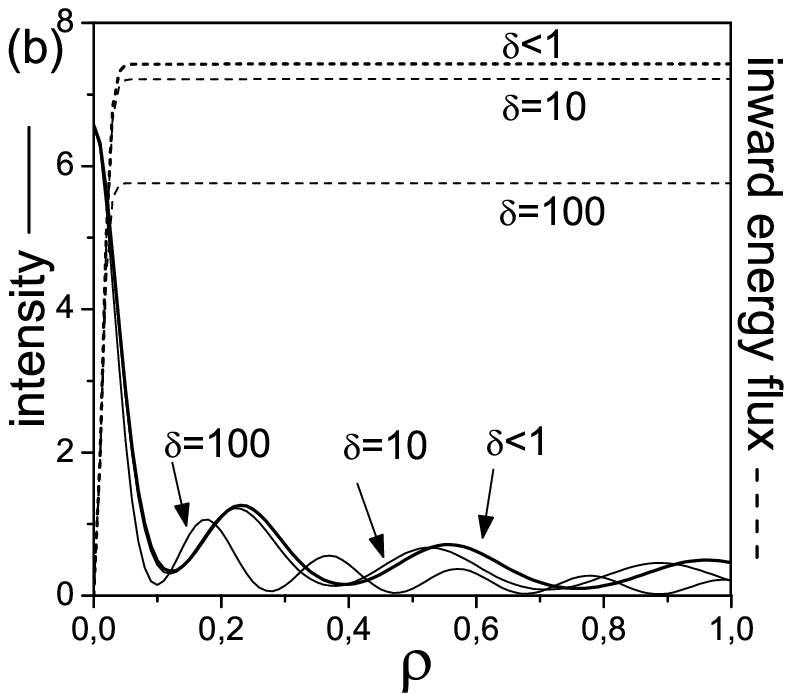}
\end{center}
\caption{\label{NL-UBB} (a) For $g=60$, $\gamma=10^{-3}$ and $K=8$, region of existence of NL-UBBs in their parameter space
$(I,\delta)$. (b) Radial intensity profiles $a^2(\rho)$ and inward radial energy fluxes $- 2\pi\rho a^2(\rho)d\phi(\rho)/d\rho$
for NL-UBBs with fixed $I=6.575$ and decreasing $\delta$ down to $0^+$, as indicated by the arrow in (a)].}
\end{figure}

Stationarity with non-negligible NLL is the distinguishing property of NL-UBBs. \cite{porras-UBB-PRL} These are solutions to the
NLSE (\ref{ANLSE}) of the form $\tilde A(\rho,\xi)=a(\rho)\exp[i\phi(\rho)-i\delta \xi]$, with $\delta>0$, whose stationary real
amplitude $a(\rho)>0$ and phase $\phi(\rho)$ must satisfy, from (\ref{ANLSE}), the ordinary differential equations
\begin{eqnarray}\label{amplitude}
\frac{1}{\rho}\frac{d}{d\rho}\left(\rho\frac{da}{d\rho}\right)-a\left(\frac{d\phi}{d\rho}\right)^2+2\delta a + 2g a^3=0\, ,\\
-2\pi\rho a^2\frac{d\phi}{d\rho} = 2\gamma 2\pi\int_0^{\rho} d\rho\rho a^{2K} \,.\label{flux}
\end{eqnarray}
Eq. (\ref{flux}) is a refilling condition, establishing that the nonlinear power loss (per unit propagation length) in any disk
of radius $\rho$ must equal to the power entering into it across its boundary (per unit length) for stationarity to be possible.
Given $g$, $\gamma$ and $K$, NL-UBBs solutions to (\ref{amplitude}) and (\ref{flux}) exist with any peak intensity $I=a^2(0)$
and axial wave number shift $\delta$ satisfying $\delta \gtrsim \mbox{min}\{0,2\eta_K \gamma I^{K-1}-gI\}$, as illustrated in
Fig. \ref{NL-UBB}(a), where $\eta_K=1.67,0.27,0.19,0.16,0.14,0.12,0.11,\dots$ for $K=2,3,\dots$. Far from its nonlinear core,
the NL-UBB approaches zero as $\rho\rightarrow\infty$ in the form of an unbalanced Bessel beam, a generalization of the
well-known Bessel beam with unequal amplitudes of its inward and outward H\"ankel beam components, and that carries, as Bessel
beams, infinite power. \cite{porras-UBB-PRL} Figure \ref{NL-UBB}(b) shows radial intensity profiles and inward radial energy
fluxes of NL-UBBs with fixed intensity $I$ and decreasing $\delta$, evaluated numerically from Eqs. (\ref{amplitude}) and
(\ref{flux}). At fixed $g$, $\gamma$ and $K$, NL-UBB profiles depend on the two parameters $I$ and $\delta$, but as $\delta$
diminishes down to $\delta=0^{+}$ with fixed $I$ [arrow in Fig. \ref{NL-UBB}(a)], the radial intensity profile of the NL-UBB is
seen to reach a limiting profile that depends only on the choice of $I$ [solid thick curve in Fig. \ref{NL-UBB}(b)].

\begin{figure}
\begin{center}
\includegraphics[width=2.93cm]{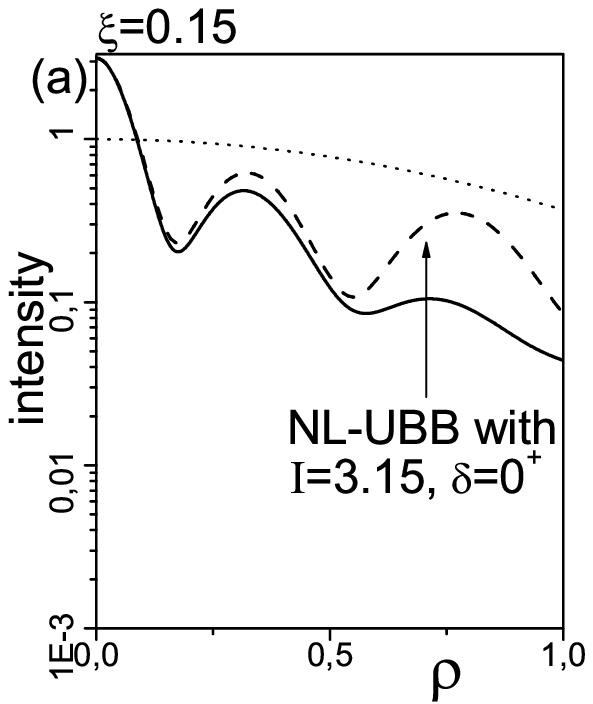}\includegraphics[width=2.75cm]{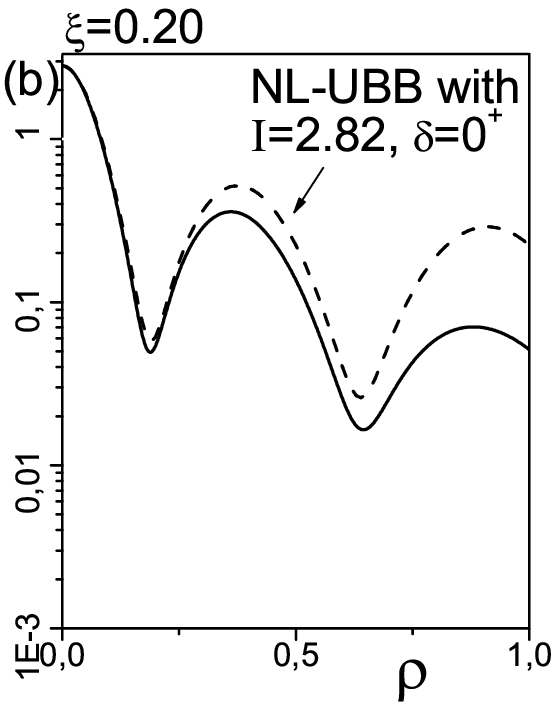}\includegraphics[width=2.78cm]{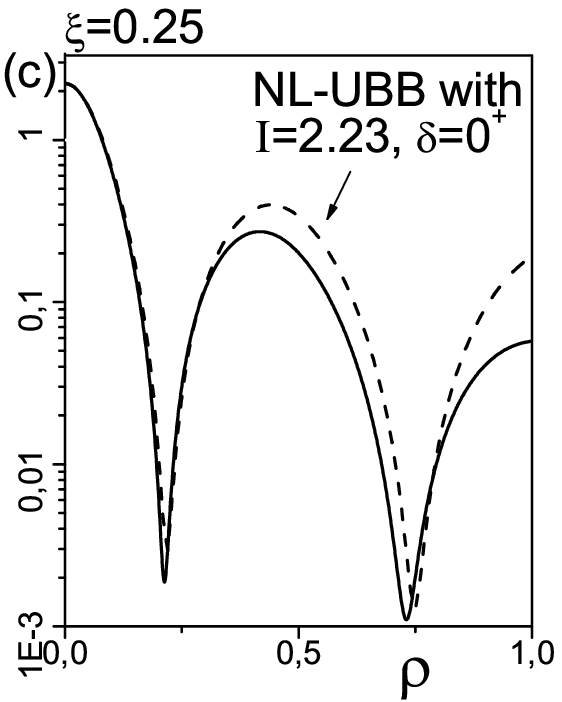}
\end{center}
\caption{\label{radial} For $g=60$, $\gamma=10^{-1}$, $K=8$ and input Gaussian (dotted curve), radial intensity profiles of the
propagated field at increasing distances beyond collapse (solid curves), and radial intensity profiles of NL-UBBs with same peak
intensities $I=3.15, 2.82,2.23$ as those of the propagated field at each distance, and $\delta= 0^+$ (dashed curves).}
\end{figure}

Coming back to the collapse of the Gaussian beam, the slow evolution along $\xi$ after collapse is interpretable as an adiabatic
sequence of NL-UBBs with slowly varying peak intensity $I$ and $\delta=0^{+}$. Figure \ref{radial} shows, for $g=60$,
$\gamma=10^{-1}$ [lowest curve in Fig. \ref{axialintensity}(a)], that the inner part of the radial intensity profile at each
propagation distance, comprising the central maximum and about one ring, fits well to that of the NL-UBBs with $\delta=0^+$ and
with $I$ equal to that of the propagated field. A truly stationary NL-UBB cannot be completely formed because of the finite
power available, which forces the slow evolution in the parameter space of NL-UBBs, with peak intensities about
$I=(g/\gamma)^{1/(K-2)}=2.9$. The fact that NL-UBBs with $\delta=0^+$ are always formed can be understood by thinking on the
collapsing beam as a bundle of rays at different angles down to zero, or equivalently, different Bessel beams with decreasing
$\delta$ (or cone angle). Each Bessel beam with $\delta\neq 0$ forms indeed its own NL-UBB with same $\delta\neq 0$ in the
nonlinear medium, as demonstrated in \cite{Pole1}, but only those with $\delta\rightarrow 0$, whose radial profile is stationary
against a change of $\delta$, emerge.

\begin{figure}[b]
\begin{center}
\includegraphics[width=4.25cm]{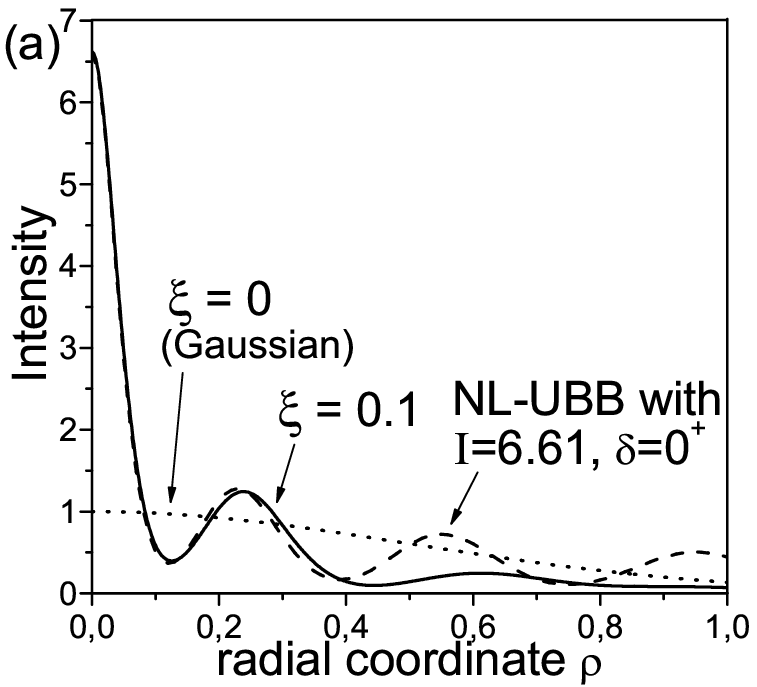}\includegraphics[width=4.25cm]{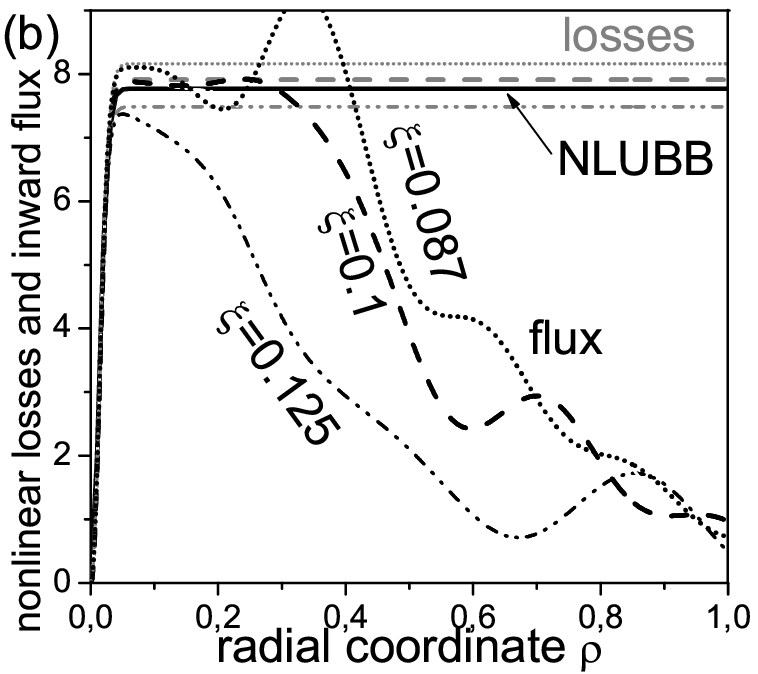}
\end{center}
\caption{\label{radial2} For $g=60$, $\gamma=10^{-3}$, $K=8$, (a) radial intensity profiles of input Gaussian, propagated field
and NL-UBB with same peak intensity $I=6.61$ as the propagated field, and $\delta = 0^+$. (b) Inward radial energy flux $F_\rho$
(black curves) and NLL $N_\rho$ (gray curves) at increasing propagation distances $\xi$. For the NL-UBB, inward flux and NLL are
equal at any $\rho$.}
\end{figure}

Figure \ref{radial2}(a) is similar to Fig. \ref{radial} but for $\gamma=10^{-3}$ with flatter peak intensity [see also Figs.
\ref{axialintensity}(a) and (b)]. The mechanism of energy refilling that sustains quasi-stationarity is evidenced in Fig.
\ref{radial2}(b). Writing $\tilde A(\rho,\xi)=a(\rho,\xi)\exp[i\varphi(\rho,\xi)]$, where $a>0$ and $\varphi$ are the real
amplitude and phase, the NLSE (\ref{ANLSE}) yields
\begin{equation}\label{energy}
\frac{d}{d\xi} 2\pi\int_0^{\rho}d\rho \rho a^2 = -2\pi\rho a^2 \partial_\rho\varphi - 2\gamma 2\pi\int_0^\rho d\rho\rho
a^{2K}\,,
\end{equation}
or $dP_\rho/d\xi=F_\rho-N_\rho$ for short, which generalizes (\ref{flux}) to a non-stationary field. Relation (\ref{energy})
expresses that the variation of the power $P_\rho$ in a disk of radius $\rho$ can be due to an inward energy flux $F_\rho$
across the disk boundary, and to the nonlinear power loss $N_\rho$ in the disk. For NL-UBBs, $dP_\rho/d\xi=0$, and
(\ref{energy}) reduces to the refilling condition (\ref{flux}), i.e., $F_\rho=N_\rho$. In particular, the total nonlinear power
loss $N_\infty$ is compensated in the NL-UBB by a constant energy flux $F_\infty$ coming from infinity [horizontal asymptota
labelled as NL-UBB in Fig. \ref{radial2}(b)], this being made feasible by the infinite power in the NL-UBB. In the collapsed
Gaussian beam, instead, the total nonlinear power loss $N_\infty$ (gray horizontal asymptotas) is not compensated by any inward
flux from infinity, since $F_\infty =0$ (black curves approaching zero), and the beam power diminishes according to
$dP_\infty/d\xi=N_\infty$ (as in Fig. \ref{power}). Nevertheless, the collapsed Gaussian beam is seen in Fig. \ref{radial2}(b)
to develop an inward energy flux in its inner portion such that the refilling condition $F_\rho= N_\rho$ characteristic of the
NL-UBBs, and hence $dP_\rho/d\xi = 0$, are approximately satisfied in the same radial region where the radial profile propagates
undistorted.

We can then conclude that NL-UBBs tend to be formed spontaneously in self-focusing media with NLL, not only from a beam with
infinite power as a Bessel beam, but also from a Gaussian beam that experiences strong self-focusing. Due to the finite power
available, the attracting NL-UBB is never reached completely, and is hence hard to be characterized. Nevertheless, the numerical
simulations indicate that the system seeks the NL-UBB with $\delta=0^+$ and a peak intensity $I\approx (g/\gamma)^{1/(K-2)}$
balancing the strengths of Kerr nonlinearity and NLL.

\end{document}